\documentclass{appolb}
\usepackage{epsfig}
\begin{document}
\title{Multiplicity fluctuations in Ar+Sc collisions at the CERN SPS from NA61/SHINE
\thanks{Presented at the Critical Point and Onset of Deconfinement 2016, Wroclaw, Poland,
May 30th - June 4th, 2016}
}
\author{Andrey Seryakov for the NA61/SHINE Collaboration
\address{St.Petersburg State University, Russia}
}
\maketitle
\begin{abstract}
{Multiplicity fluctuations were investigated in Ar+Sc collisions at 13A, 19A, 30A, 40A, 75A, 150A GeV/c using the NA61/SHINE detector at the SPS. Centrality selection is based on the nucleon-spectator energy in the forward hemisphere as measured by the Projectile spectator detector. Preliminary results on the scaled variance $\omega$ and the strongly intensive measure $\Omega$ were obtained for the multiplicity distribution of all, negatively and positively charged hadrons. Results are discussed and compared with p+p results and Pb+Pb data of NA49 and EPOS 1.99 simulations.}
\end{abstract}
\section{Centrality determination in NA61/SHINE}
Centrality is a key parameter in heavy ion collisions as it allows to restrict the volume of the created system. The classical definition of centrality is based on impact parameter which is not directly measurable. Therefore experiments use different techniques to define centrality indirectly. The most common method is to apply restrictions on the produced particle multiplicity in some pseudorapidity intervals \cite{ALICE centrality}. This definition possibly biases multiplicity fluctuation measurements. Therefore NA61/SHINE uses for centrality determination information from a forward hadron colorimeter PSD \cite{Abgrall:2014fa}, which measures the energy of all non-interacted forward nucleons. 
\section{Fluctuation measures}
Fluctuation quantities are studied to probe the critical point CP of strongly interacting matter. The most widely used is the scaled variance of the multiplicity distribution:
\begin{eqnarray}
            &\omega[N] = (\langle N^{2}\rangle -{\langle N\rangle}^{2})/\langle N\rangle,
\end{eqnarray}
where $\langle\cdots\rangle$ stands for averaging over all events. Within the Wounded Nucleon Model (WNM) or the ideal Boltzmann multi-component gas in the grand canonical ensemble (IB-GCE) one can write \cite{Gorenstein:2011vq}:
\begin{eqnarray}
            &\omega[N] = \omega[n]+\omega[W]\langle N\rangle/\langle W\rangle,
\end{eqnarray}
where $n$ is the multiplicity produced from one wounded nucleon or from a fixed volume (IB-GCE) and $W$ is the number of wounded nucleons. To probe the CP it is important to suppress the volume fluctuation ($\omega[W]$). Therefore the scaled variance should be measured for the most central collisions or strongly intensive quantities should be used. In Refs.~\cite{Gorenstein:2011vq,Gazdzicki:2013ana} the following measures were proposed:
\begin{eqnarray}
            &\Delta[A,B] = \frac{1}{C_{\Delta}} \biggl[ \langle B \rangle \omega[A] -
                        \langle A \rangle \omega[B] \biggr] \\
            &\Sigma[A,B] = \frac{1}{C_{\Sigma}} \biggl[ \langle B \rangle \omega[A] +
                        \langle A \rangle \omega[B] - 2 \bigl( \langle AB \rangle -
                        \langle A \rangle \langle B \rangle \bigr) \biggr],
\end{eqnarray}
where $A$ and $B$ are extensive quantities and $C_{\Delta}$ and $C_{\Sigma}$ are normalization coefficients, which are chosen so that $\Delta=\Sigma=1$ for the models of independent particle production.
A new quantity $\Omega$~\cite{Poberezhnyuk:2015Omega} can be constructed from $\Delta$ and $\Sigma$ by setting $C_{\Delta}=C_{\Sigma}=\langle B \rangle$:
\begin{eqnarray}
&\Omega[A,B]=\frac{1}{2}(\Delta[A,B]+\Sigma[A,B]),
\end{eqnarray}
If $A$ and $B$ are uncorrelated from a fixed volume within the IB-GCE or from a single source then $\Omega[A,B]=\omega[a]$, where $\omega[a]$ is $\omega[A]$ in the fixed volume or the scaled variance of $A$ from a single source. Therefore for the most central collisions one expects that $\Omega[A,B]\approx\omega[A]$.
\section{Analysis procedure}
\begin{figure}[h]
\begin{minipage}{10pc}
\includegraphics[width=9pc]{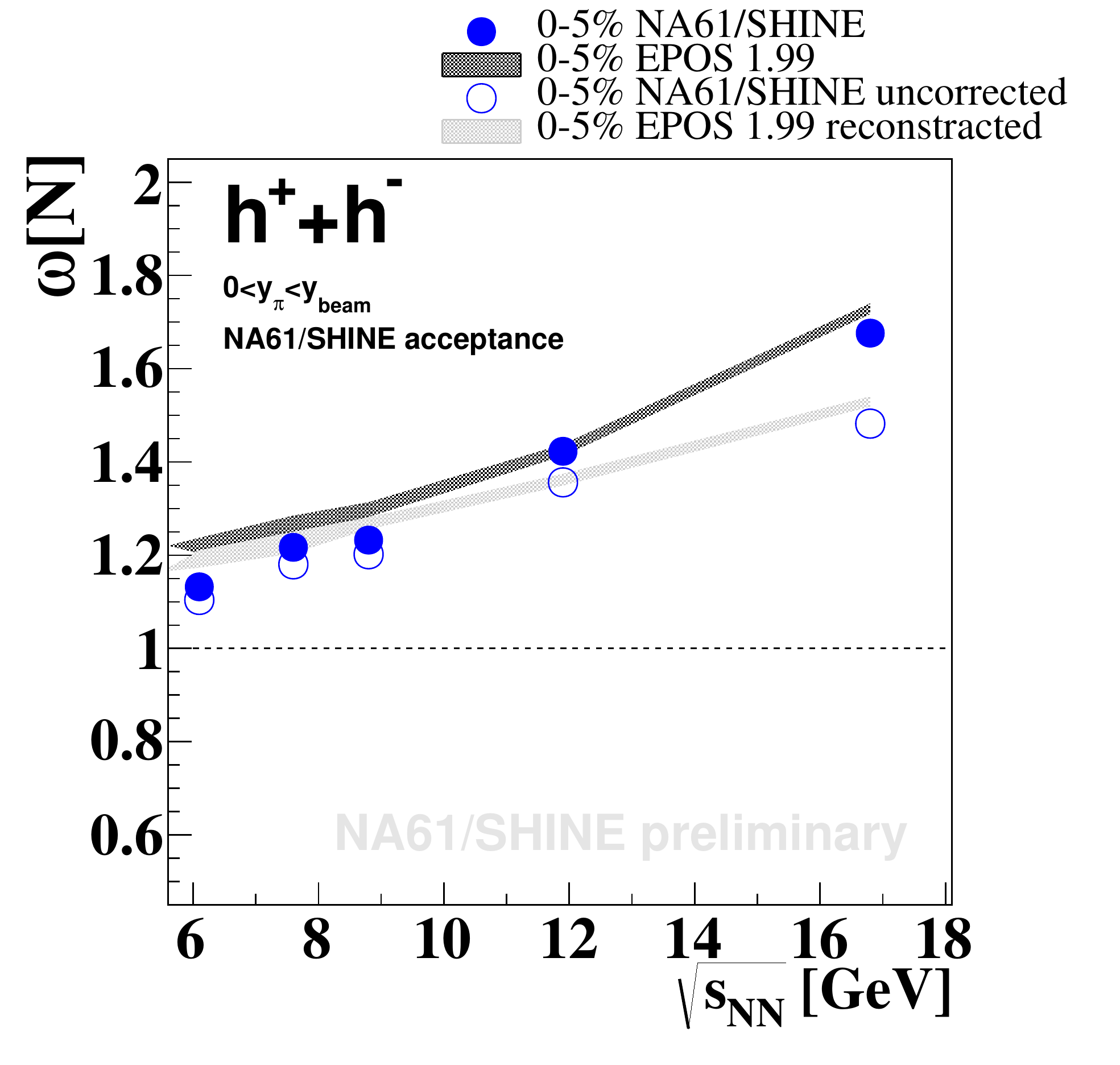}
\end{minipage}
\begin{minipage}{10pc}
\includegraphics[width=9pc]{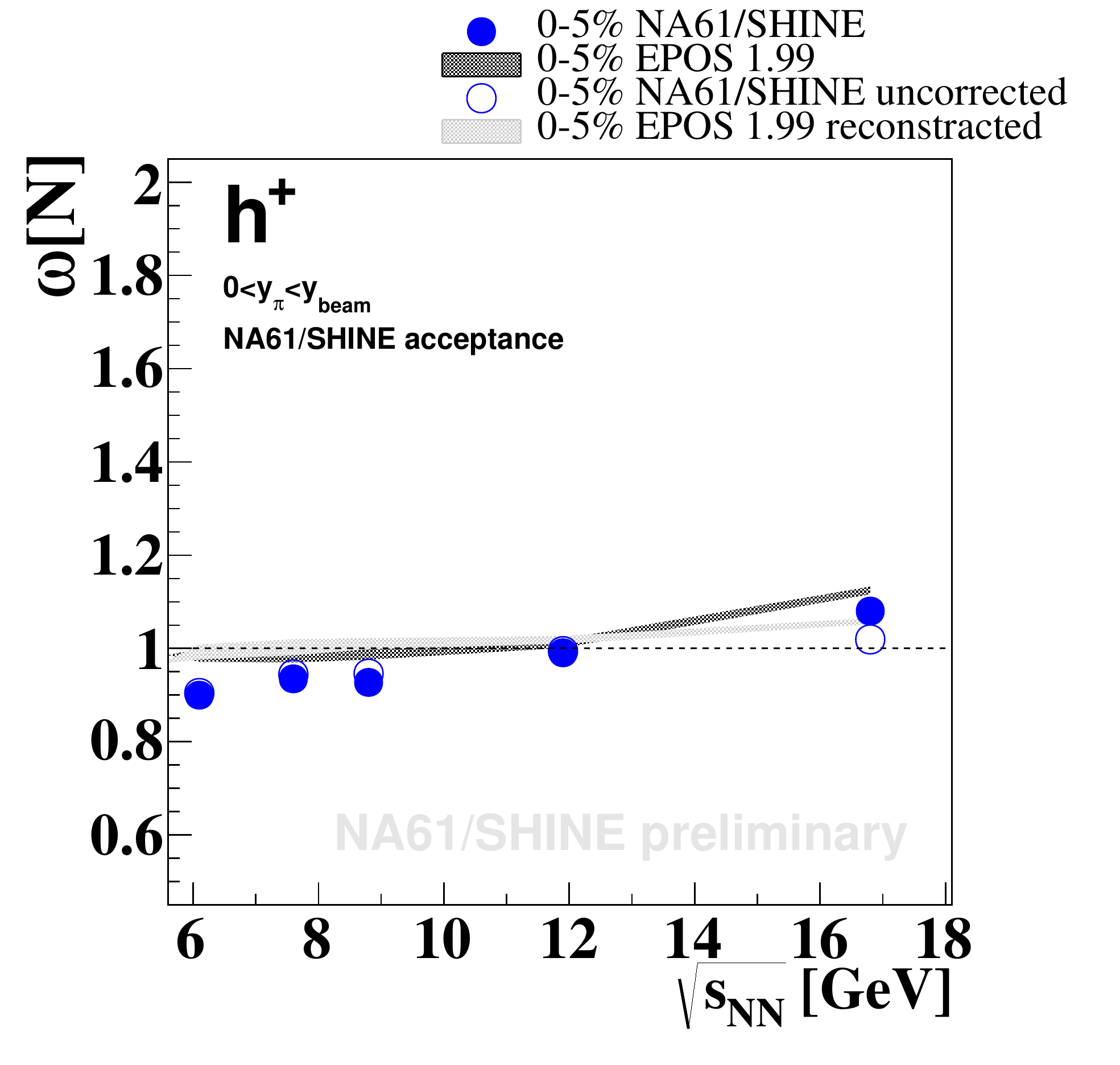}
\end{minipage}
\begin{minipage}{10pc}
\includegraphics[width=9pc]{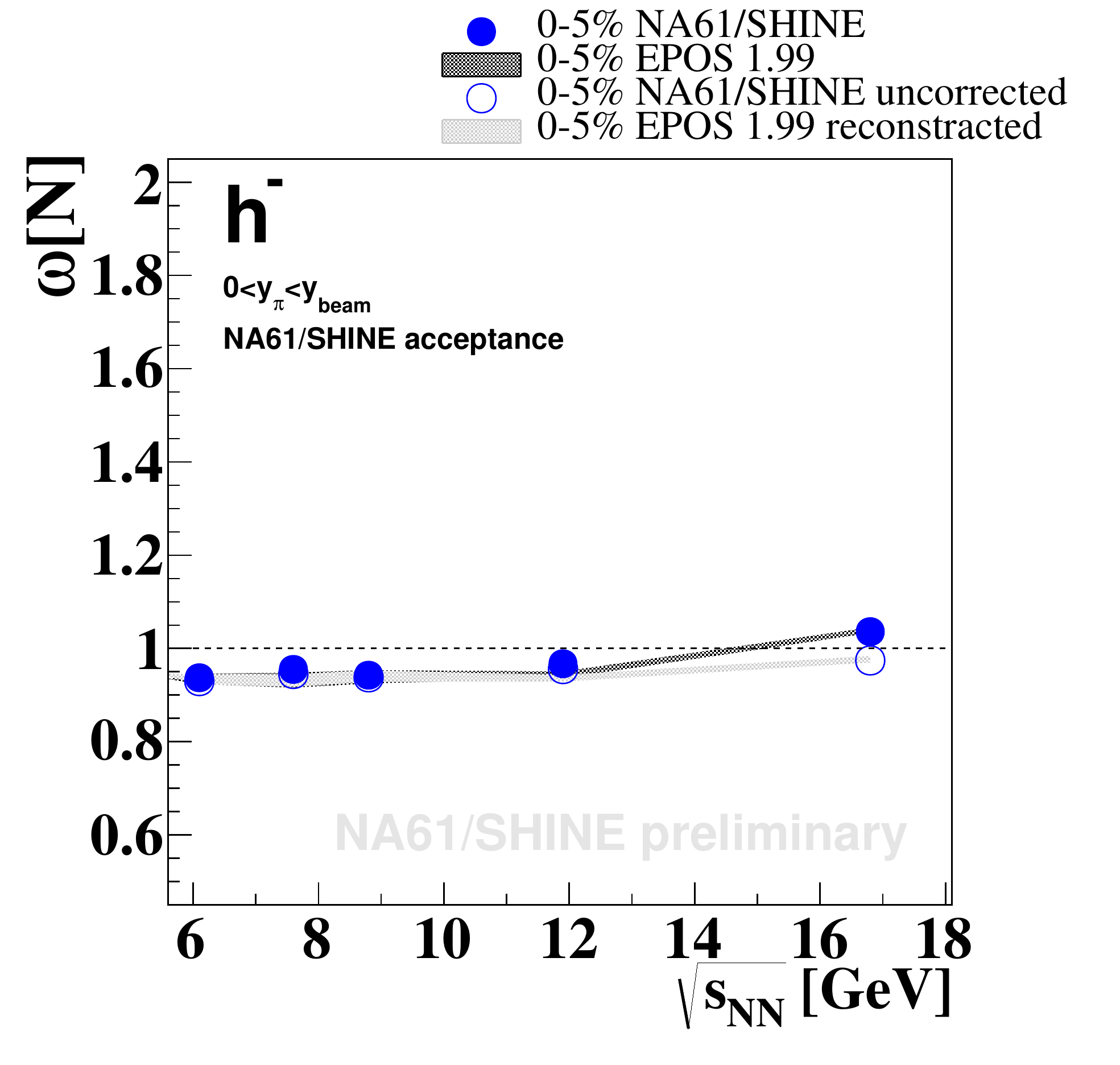}
\end{minipage}
\caption{\label{label}The effect of correction procedure on $\omega[N]$ of charged hadrons for the 0-5$\%$ forward energy selected Ar+Sc data. The open dots are uncorrected data, the full dots are corrected, the black line is the pure EPOS 1.99 simulation, the grey line shows the EPOS 1.99 results after detector simulation and reconstruction. Results are for $0<y_{\pi}<y_{beam}$ and the NA61/SHINE acceptance.}
\label{corrections}
\end{figure}

Preliminary results were obtained from forward energy selected Ar+Sc collisions at 19, 30, 40, 75 and 150{\it A} GeV/c respectively. The analysis was performed in the NA61/SHINE acceptance with the restriction $0<y^{*}_{\pi}<y_{beam}$, where $y^{*}$ is the particle rapidity in the c.m.s. under pion mass assumption. The rapidity cut was introduced to exclude regions of poor azimuthal angle acceptance and electron contamination in backward hemisphere and to reduce diffractive effects in the forward region. Transverse momenta of all charged particles were restricted to $0<p_{T}<1.5$ GeV/c. 
Event and track selection criteria were chosen to select only inelastic interactions and hadrons produced in strong and EM processes. The p+p data~\cite{Czopowicz:na61} shown for comparison were corrected for contributions from non-target interactions, for Ar+Sc collisions these corrections were estimated to be negligible. The results are corrected for detector inefficiencies and trigger biases by multidimensional moments corrections for p+p collisions\cite{Czopowicz:na61} and by separate moments corrections for Ar+Sc collisions. The effect of the corrections on $\omega[N]$ for Ar+Sc is shown  in Fig.~\ref{corrections}. $\Omega[N,E_{P}]$ is presented in this paper without corrections.
Statistical uncertainties were calculated using the sub-sample method. Systematic uncertanties were estimated to be less than 10$\%$, further investigations will follow. The systematic uncertainties are not shown in the figures.  
The centrality determination procedure in EPOS 1.99 is based on the number of forward nucleon-spectators for correct comparison between data and simulation.
\section{Results}
\begin{figure}[h]
\begin{minipage}{10pc}
\includegraphics[width=9pc]{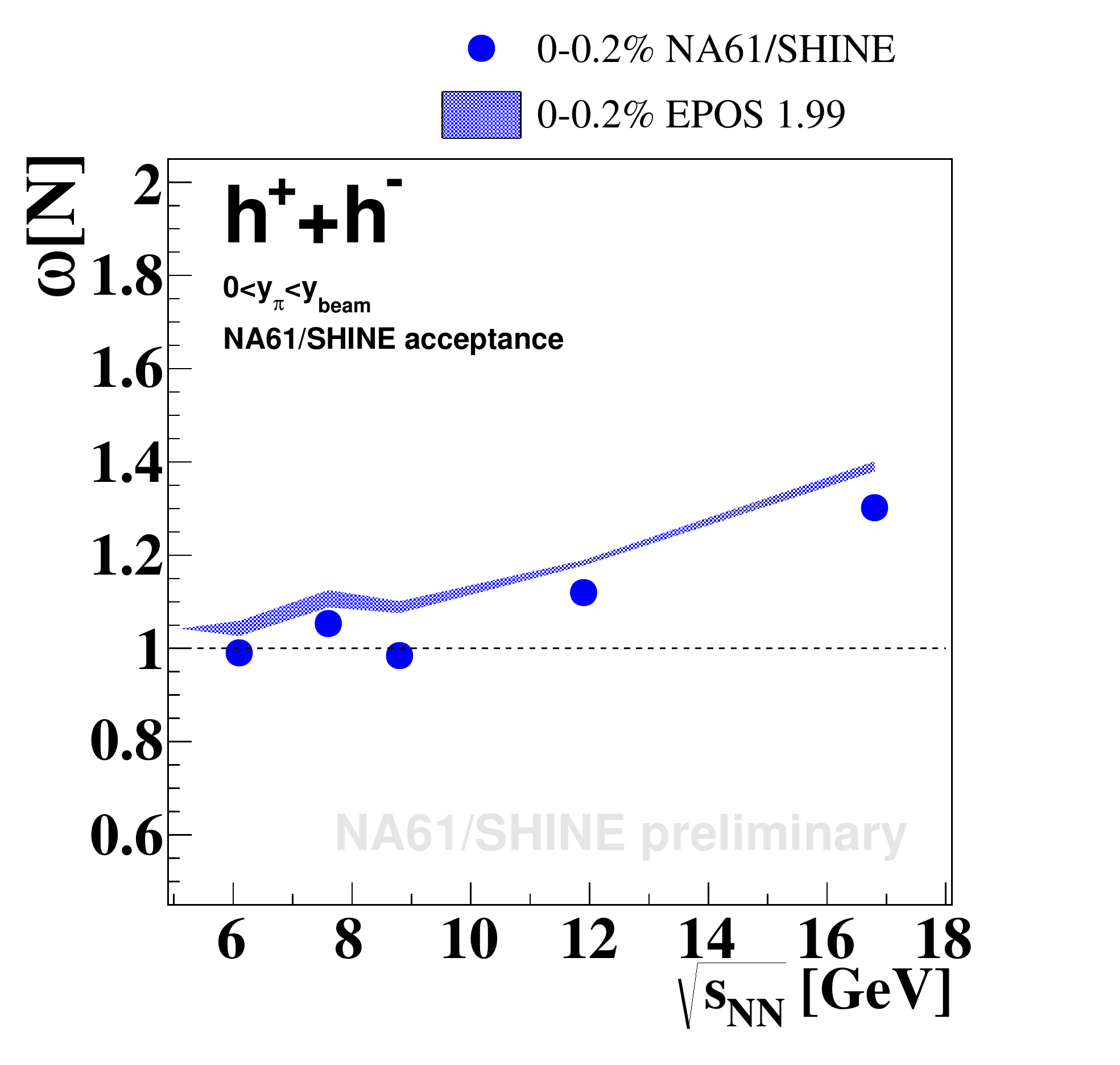}
\end{minipage}
\begin{minipage}{10pc}
\includegraphics[width=9pc]{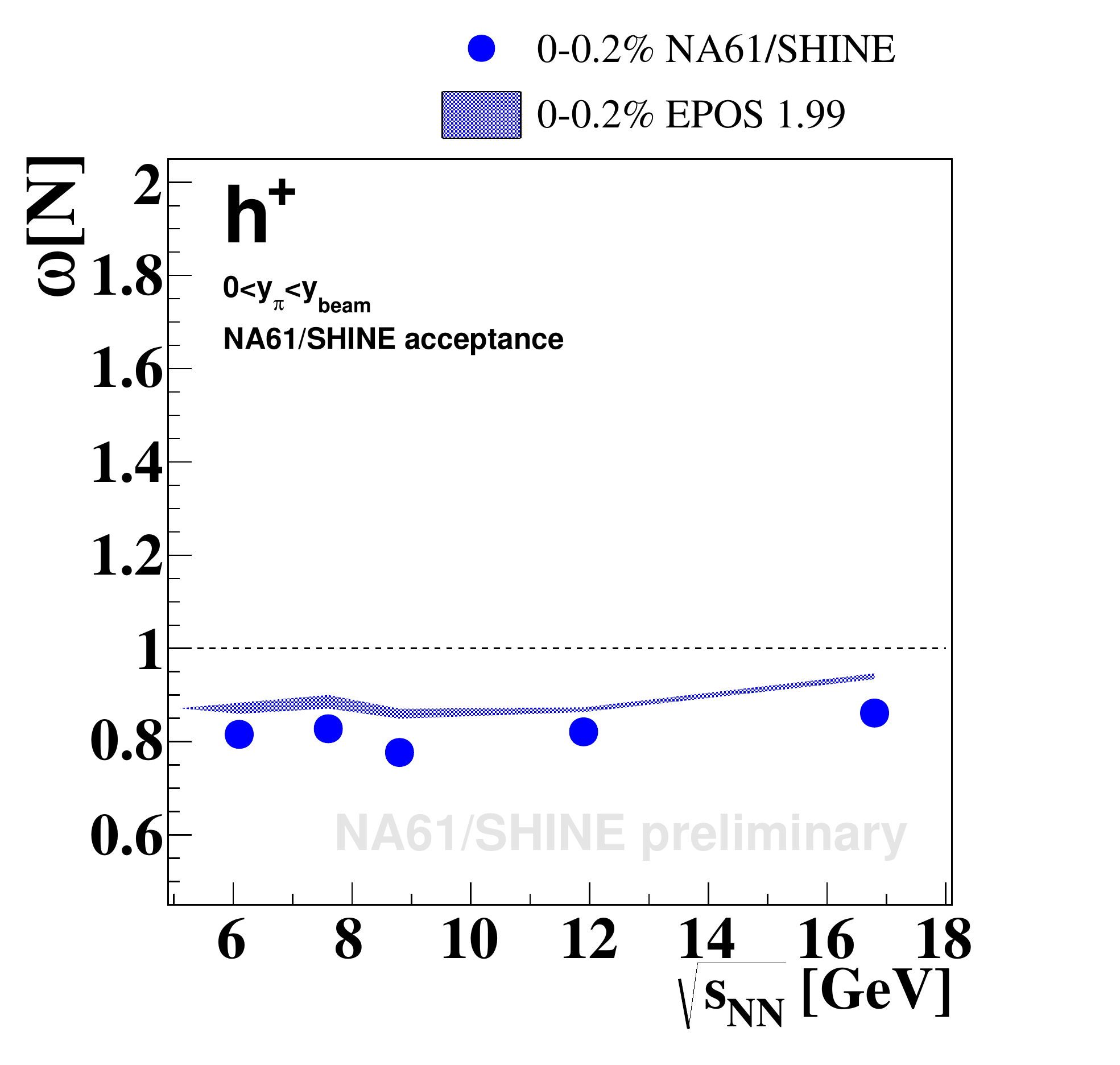}
\end{minipage}
\begin{minipage}{10pc}
\includegraphics[width=9pc]{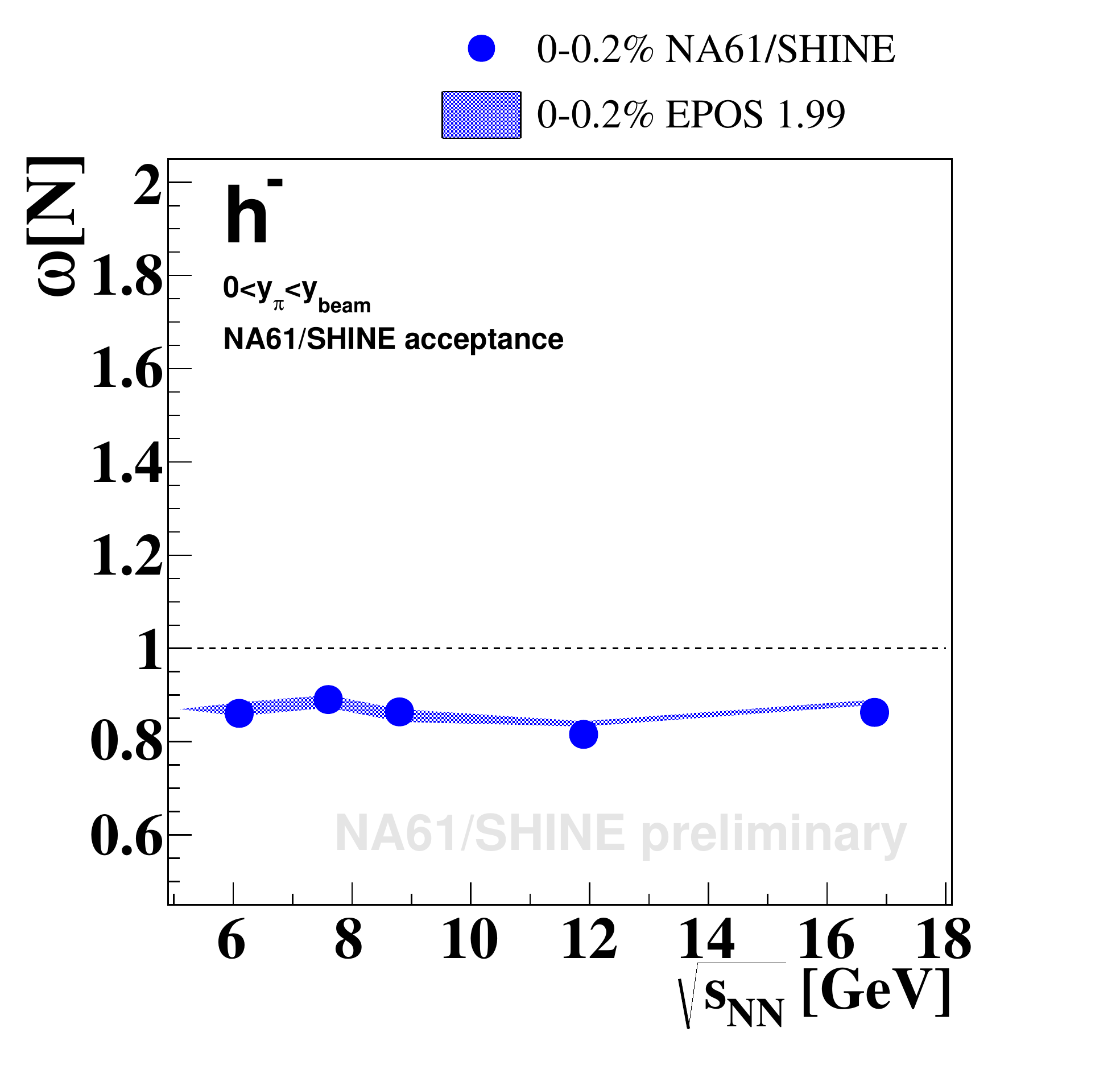}
\end{minipage}
\caption{\label{label} The energy dependence of $\omega[N]$ for the 0-0.2$\%$ most central Ar+Sc collisions ins the NA61/SHINE acceptance and $0<p_{T}<1.5$ GeV/c and $0<y_{\pi}<y_{beam}$. EPOS 1.99 predictions are shown the blue line.}
\label{EPOS}
\end{figure}

\begin{figure}[h]
\begin{minipage}{10pc}
\includegraphics[width=9pc]{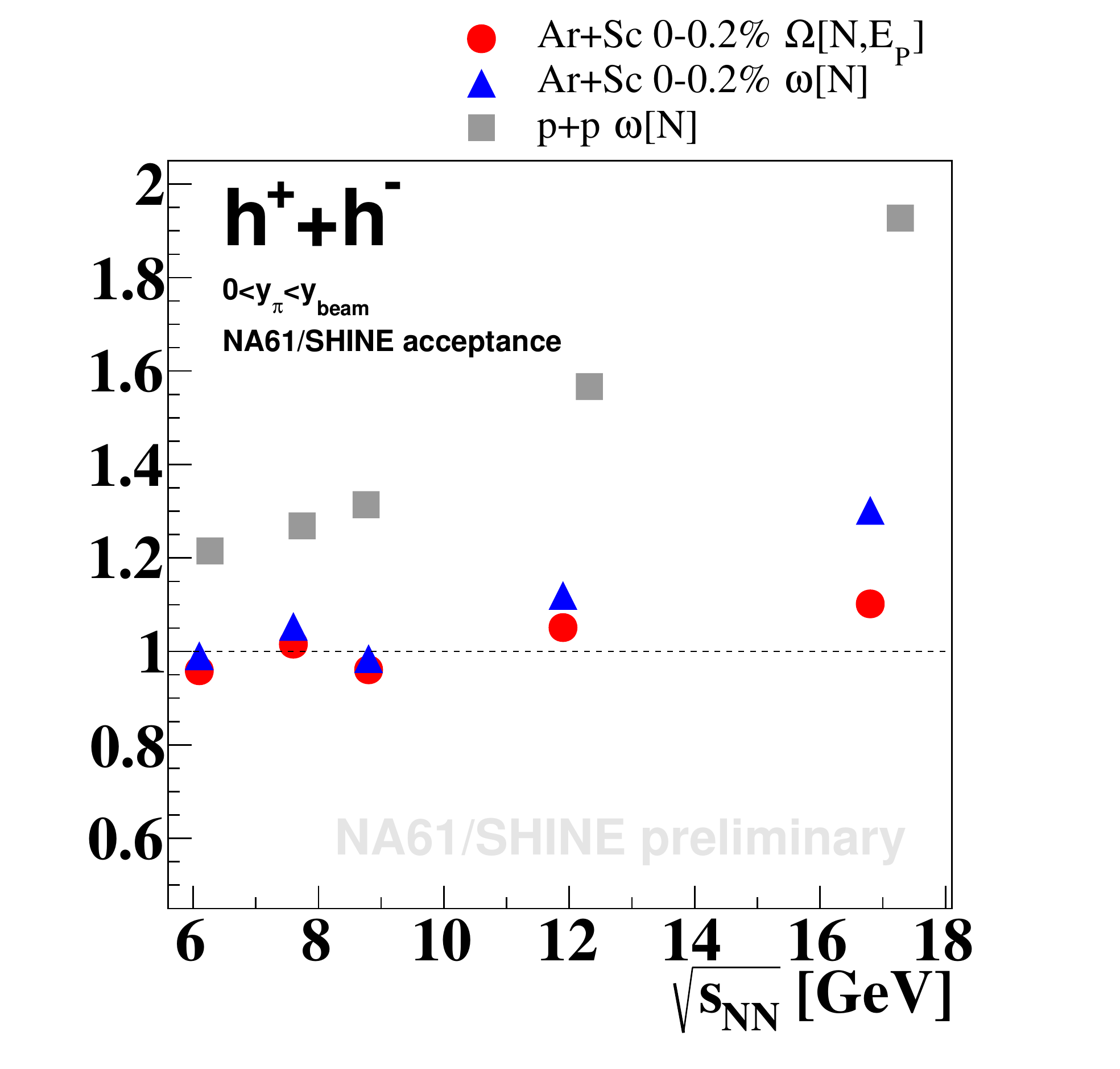}
\end{minipage}
\begin{minipage}{10pc}
\includegraphics[width=9pc]{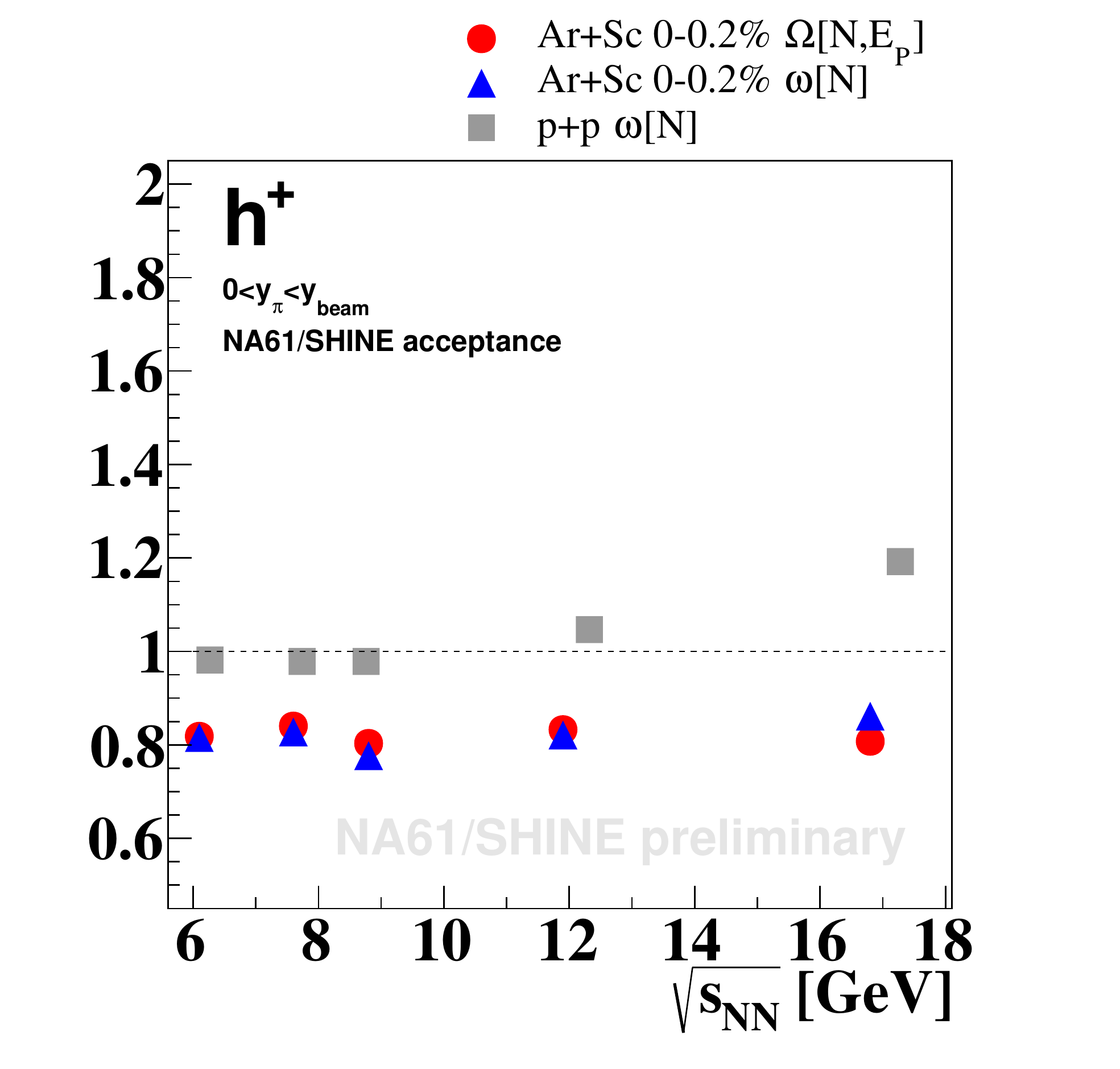}
\end{minipage}
\begin{minipage}{10pc}
\includegraphics[width=9pc]{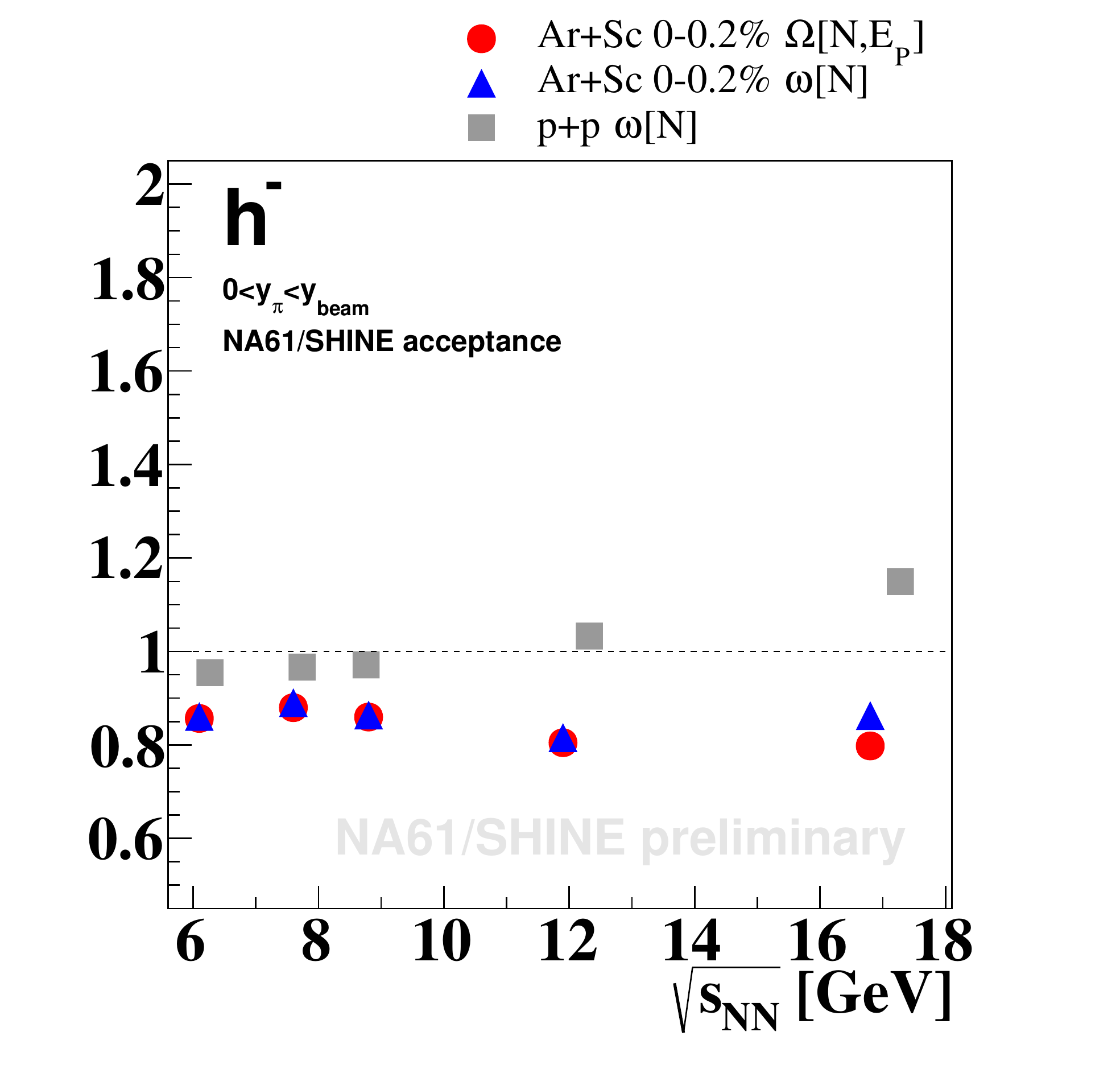}
\end{minipage}
\caption{\label{label}The energy dependences of $\Omega[N,E_{P}]$ (red dots) and $\omega[N]$ (blue triangles) for 0-0.2$\%$ central Ar+Sc collisions compared with $\omega[N]$ for p+p data (grey squares). Results are inside the NA61/SHINE acceptance and for $0<y_{\pi}<y_{beam}$ and $0<p_{T}<1.5$ GeV/c.}
\label{Omega}
\end{figure}

\begin{figure}[h]
\centering
\includegraphics[width=9pc]{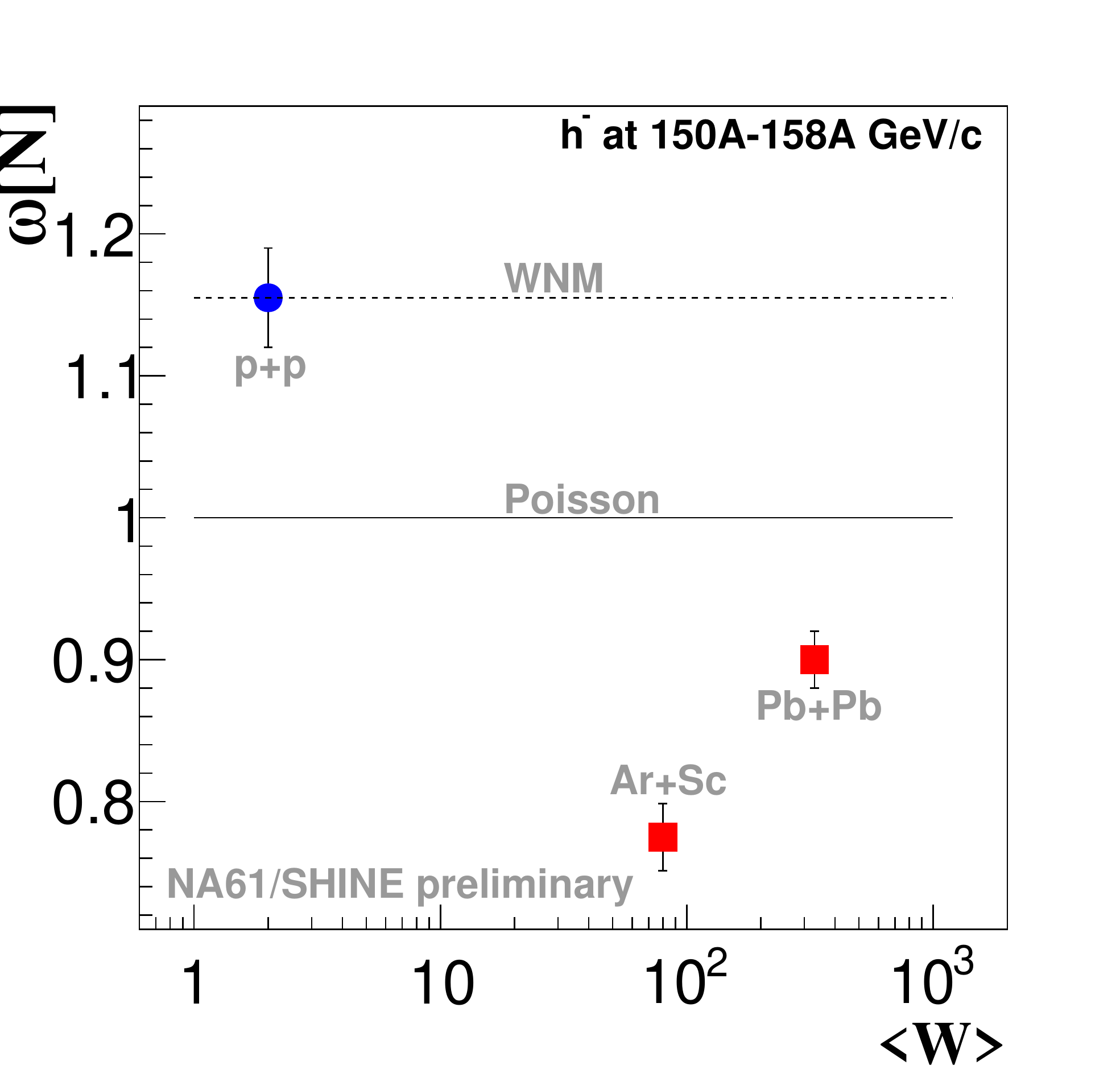}
\caption{\label{label}Scaled variance ($\omega[N]$) for negatively charged hadrons measured in p+p (circle)\cite{Czopowicz:na61} as well as 0-1\% Pb+Pb \cite{Alt:2007jq}, and 0-0.2\% Ar+Sc collisions at 150/158{\it A} GeV/c (squares). Results are for $0< y_{\pi}< y_{beam}$ and the NA49B acceptance \cite{Alt:2007jq}. Experimental data are compared to predictions of the Wounded Nucleon Model as well as to Statistical Models.}
\label{wounded}
\end{figure}

The energy dependence of the scaled variance of the multiplicity of all, positively and negatively charged hadrons ($\omega[N]$) is shown in Figs.~\ref{corrections}~and~\ref{EPOS} for 0-5$\%$ and 0-0.2$\%$ forward energy selected Ar+Sc collisions. A small enhancement at 8{\it A} GeV/c is observed for the most central collisions. As EPOS1.99 simulations show the same behaviour it probably is an acceptance effect. 
The charged hadron multiplicity and the difference between the beam energy and the forward energy measured by the PSD($E_{P}=E_{beam}-E_{PSD}$) were chosen as variables for the $\Omega$ quantity as these values are uncorrelated from the singe source for the independent particle production model. With this assumption $\Omega[N,E_{P}]=\omega[n]\approx\omega[N]$ for the most central collisions.
The energy dependence of $\Omega[N,E_{P}]$ and $\omega[N]$ for the 0.2$\%$ most central Ar+Sc collisions are shown in the Fig.~\ref{Omega} and compared with $\omega[N]$ for the p+p data. 
There is no significant non-monotonic behaviour in the energy dependence of those two measures, however an interesting effect was observed (see Fig.~\ref{wounded}). Namely, for negatively charged hadrons (the cleanest sample) at 150/158{\it A} GeV/c the scaled variance of the multiplicity distribution is much below 1 for 0-0.2\% Ar+Sc and 0-1$\%$ Pb+Pb and above 1 for p+p collisins. Within the Wounded Nucleon Model (WNM) one expects that $\omega[N]_{AA}$ should be grater or equal $\omega[N]_{pp}$. Therefore, the result obtained by NA61/SHINE and NA49 clearly shows the violation of the Wounded Nucleon Model in these reactions.
In the IB-GCE the multiplicity distribution is Poissonian ($\omega[N]=1$), independent of the (fixed) system volume, and thus $\omega[N]_{AA}<1$ is forbidden. On the other hand, $\omega[N]$ can increase due to resonance decays and Bose-Einstein statistics and decrease due to conservation laws \cite{Begun:2006uu}. In fact, the NA49 Pb+Pb point is very well described by the hadron gas model in the micro canonical ensemble (HG-MCE) \cite{Begun:2006uu}. Within the statistical models the result $\omega[N] \gg 1$, as seen in p+p, can be understood as a result of volume and/or energy fluctuations \cite{Begun:2008fm}.
{\small \textbf{Acknowledgments:} This work was supported by the SPbSU research grant 11.38.242.2015.}

\end{document}